\documentclass[a4paper]{jpconf}
\usepackage[dvipdfmx]{graphicx}
\begin{document}
\title{Counterflow quantum turbulence in a square channel under the normal fluid with a Poiseuille flow}

\author{Satoshi Yui$^1$ and Makoto Tsubota$^{1,2}$}

\address{$^1$ Department of Physics, Osaka City University, 3-3-138 Sugimoto, Sumiyoshi-Ku, Osaka 558-8585, Japan}
\address{$^2$ The OCU Advanced Research Institute for Natural Science and Technology (OCARINA), Osaka City University, 3-3-138 Sugimoto, Sumiyoshi-Ku, Osaka 558-8585, Japan}

\ead{yui@sci.osaka-cu.ac.jp}

\begin{abstract}
  We perform a numerical analysis of superfluid turbulence produced by thermal counterflow in He II by using the vortex filament model.
  Counterflow in a low aspect ratio channel is known to show the transition from laminar flow to the two turbulent states TI and TII.
  The present understanding is that the TI has the turbulent superfluid and the laminar normal fluid but both fluids are turbulent in the TII state.
  This work studies the vortex tangle in the TI state.
  Solid boundary condition is applied to walls of a square channel, and the velocity field of the normal fluid is prescribed to be a laminar Poiseuille profile.
  An inhomogeneous vortex tangle, which concentrates near the solid boundaries, is obtained as the statistically steady state.
  It is sustained by its characteristic space-time oscillation.
  The inhomogeneity of the vortex tangle shows the characteristic dependence on temperature, which is caused by two competitive effects, namely the profile of the counterflow velocity and the mutual friction.
\end{abstract}

\section{Introduction}
  A thermal counterflow in He II is internal convection of two fluids, namely the normal fluid and the superfluid.
  When the counterflow velocity exceeds a critical value, a self-sustaining tangle of quantized vortices appears to form superfluid turbulence.
  In low aspect ratio channels, superfluid turbulence makes the mysterious transition.
  The increase in the counterflow velocity is observed to change the laminar state to the first turbulent state TI, and next to the second turbulent state TII \cite{tough}.
  Melotte and Barenghi suggested that the transition from the TI to TII state is caused by the transition of the normal fluid from laminar to turbulent \cite{melotte}.
  The recent developments of the visualization technique have enabled us to confirm the scenario.
  Guo $\it et ~  al.$ have followed the motion of seeded metastable ${\rm He}_2^*$ molecules by a laser-induced-fluoresence technique to observe that the normal fluid can be turbulent at relatively high velocities \cite{guo}.

  To understand the mysterious transition of counterflow quantum turbulence, it is necessary to address the coupled dynamics of the two fluids;
  the superfluid is described by the vortex filament model, the normal fluid by the Navier-Stokes equation, and they are coupled through the mutual friction \cite{kivotides}.
  However, it is difficult to solve fully the coupled dynamics.
  As the first essential step, we address the TI state in a square channel with prescribing the velocity field of the normal fluid to a Poiseuille profile.
  Our simulation obtains a statically steady state of an inhomogeneous vortex tangle.

  Baggaley $\it et ~  al.$ \cite{baggaley} studied numerically a thermal counterflow between two plates.
  They prescribed a Poiseuille and turbulent profiles for the velocity field of the normal fluid.
  An inhomogeneous vortex tangle was obtained, where vortices concentrated near the solid boundaries.
  They suggested that their results supported the scenario proposed by Melotte and Barenghi.
  The better understanding of the TI and TII states would be obtained by studying the flow in a low aspect ratio channel where all boundaries are solid except for the flow direction.
  This is because the TI and TII states are actually observed in low aspect ratio channels and another turbulence TIII state is observed in high aspect ratio channels \cite{tough}.

\section{Formulation}
  In a vortex filament model \cite{schwarz85} a quantized vortex is represented by a filament passing through a fluid and has a definite vorticity.
  This approximation is very suitable in He II, since the core size of a quantized vortex is much smaller than any other characteristic length scale.
  At zero temperature the vortex filament moves with the superfluid velocity
  $
    {\bf v}_s = 
        {\bf v}_{s,\omega}
      + {\bf v}_{s,b}
      + {\bf v}_{s,a}
      ,
  $
  where ${\bf v}_{s,\omega}$ is the velocity field produced by vortex filaments, ${\bf v}_{s,b}$ by solid boundaries, and ${\bf v}_{s,a}$ is the applied superfluid velocity.
  The velocity field ${\bf v}_{s,\omega}$ is given by the Biot-Savart law;
  this work addresses the full Biot-Savart integral \cite{adachi}.
  Since ${\bf v}_{s,a}$ represents the laminar flow of the superfluid, ${\bf v}_{s,a}$ requires irrotational condition, which is supposed to be uniform.
  The velocity field ${\bf v}_{s,b}$ is obtained by a simple procedure;
  it is just the field produced by an image vortex which is constructed by reflecting the filament into the surface and reversing its direction.
  Taking into account the mutual friction, the velocity of a point $\bf s$ on the filament is given by
  \begin{equation}
    {\dot{\bf s}} = {{\bf v}_s}
      + \alpha {\bf s}' \times ({\bf v}_n - {{\bf v}}_s)
      - \alpha '{\bf s}' \times [{\bf s}'\times({\bf v}_n - {{\bf v}}_s)],
  \end{equation}
  where $\alpha$ and $\alpha'$ are the temperature-dependent coefficients, and the prime denotes derivatives of ${\bf s}$ with respect to the coordinate $\xi$ along the filament.
  The velocity field of the normal fluid is prescribed to be a Poiseuille profile.
  In a rectangular channel the Poiseuille profile is given by
  \begin{equation}
    v_{n} = u_{0} \sum _{m=1,3,5,\cdots} ^{\infty} (-1)^{\frac{m-1}{2}}
      \left[
        1 - \frac{ \cosh(m \pi z / 2 a) }{ \cosh(m \pi b / 2 a) }
      \right]
      \frac{ \cos(m \pi y / 2 a) }{ m^3 }
      ,
  \end{equation}
  where $y$ and $z$ are coordinates normal to the flow direction $x$, and $a$ and $b$ are halves of the channel width along the $y$- and $z$- axes \cite{poiseuille}.

\section{Numerical simulation}
  In this study, all simulations are performed under the following conditions.
  We study thermal counterflow of He II at temperatures {\it T}=1.3~K, 1.6~K and 1.9~K.
  The computing box is $0.1 \times 0.1 \times 0.1 \mathrm{~cm^3}$.
  Periodic boundary conditions are used along the flow direction $x$, while solid boundary conditions are applied to the channel walls at $y=\pm a$ and $z=\pm b$.
  All simulations start with eight randomly oriented vortex rings of radius $0.023 \mathrm{~cm}$.

  The vortex line density (VLD) is defined as $L = \frac{1}{\Omega} \int _{\cal L} d\xi$, where the integral is performed along all vortices in the sample volume $\Omega$.
  The vortex tangle reaches the statistically steady state.
  Figure 1 (a) shows the time evolution of VLD.
  Fluctuations are larger than those in a uniform counterflow \cite{adachi}, which is attributable to the mechanism characteristic of this system discussed in section 4.1.

  \begin{figure}[t]
    \begin{center}
    \begin{minipage}[b]{0.40\textwidth}
      \includegraphics[width=1.0\textwidth]{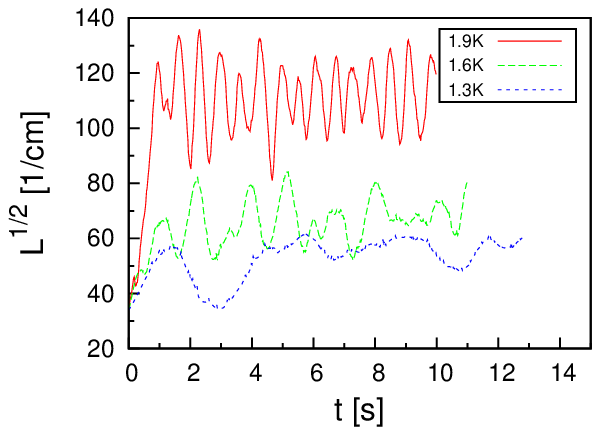}
      (a)
    \end{minipage}
    \hspace{0.01\textwidth}
    \begin{minipage}[b]{0.40\textwidth}
      \includegraphics[width=1.0\textwidth]{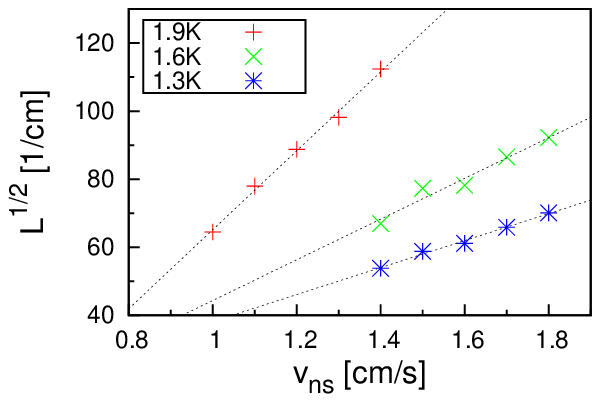}
      (b)
    \end{minipage}
    \end{center}
    \caption
    {
      (Color online) 
      (a) Vortex line density as a function of time with $v_{ns}=1.4 \mathrm{~cm/s}$.
      (b) Vortex line density averaged over the statistically steady state as a function of $v_{ns}$.
    }
    \label{dens.eps}
  \end{figure}

  The statistically steady state is known to exhibit the characteristic relation $L^{1/2} = \gamma (v_{ns} - v_{0})$ \cite{vinen} with the parameters $\gamma$ and $v_{0}$.
  We regard the counterflow velocity $v_{ns}$ as the spatially averaged amplitude of ${\bf v}_{ns}={\bf v}_{n}-{\bf v}_{s,a}$.
  Figure 1 (b) shows the VLD temporally averaged over the statistically steady state, which almost satisfies the relation.
  Table 1 shows the comparison of $\gamma$ among the present work $\gamma_{\rm num1}$, the simulation $\gamma_{\rm num2}$ under the periodic boundary condition \cite{adachi} and a typical experiment $\gamma_{\rm exp}$ \cite{childers}.
  The values of $\gamma_{\rm num1}$ are lower than the values of $\gamma_{\rm num2}$ obtained under the uniform counterflow. 
  The difference of $\gamma$ comes from the difference of the mechanism sustaining the vortex tangle.
  The origin of the discrepancy between $\gamma_{\rm num1}$ and $\gamma_{\rm exp}$ is not clear, but this may be attributable to neglecting the effect of the vortex tangle on the Poiseuille flow of the normal fluid through mutual friction.

  \begin{table}[b]
  \begin{tabular}{cccc}
    \hline
    $T$& $\gamma_{\rm num1}$ & $\gamma_{\rm num2}$ & $\gamma_{\rm exp}$ \\
    (K)& (s/cm$^2$) &(s/cm$^2$)  &(s/cm$^2$) \\ \hline
            1.3 & 40    & 53.5  &   59   \\
            1.6 & 60    & 109.6 &   93   \\
            1.9 & 112   & 140.1 &   133  \\
    \hline
  \end{tabular}
  \hspace{0.02\textwidth}
  \begin{minipage}[b]{0.57\textwidth}
    \caption
    {
      Line density coefficients $\gamma$.
      Numerical results $\gamma_{\rm num1}$ by this work, $\gamma_{\rm num2}$ by Adachi {\it et al.} \cite{adachi} and experimental results $\gamma_{\rm exp}$ by Childers and Tough \cite{childers}.
    }
    \label{gamma}
  \end{minipage}
  \end{table}

  In order to estimate the inhomogeneity of the vortex tangle, we divide the computational box to $15^3$ sub-volumes and define the VLD at a sub-volume as the local VLD.
  Figure 2 (a) shows the spatially dependence of $L'(y,z)$, which is obtained by averaging the local VLD spatially over the flow direction and temporally over the statistically steady state.
  We estimate the inhomogeneity of the vortex tangle by a spatial variance ${\it \Delta}_L$ of $L'/L$.
  Figure 2 (b) shows the characteristic dependence of ${\it \Delta}_L$ on $v_{ns}$ and $T$.
  Firstly, ${\it \Delta}_L$ of 1.6K is the largest among three temperatures.
  Secondly, ${\it \Delta}_L$ at 1.3K and 1.6K increases with $v_{ns}$, while ${\it \Delta}_L$ at 1.9~K decreases with $v_{ns}$.
  The dependence on $T$ is understood as discussed in section 4.2, but the dependence on $v_{ns}$ is not known.

  \begin{figure}[t]
  \begin{center}
  \begin{minipage}[b]{0.40\textwidth}
    \includegraphics[width=1.0\textwidth]{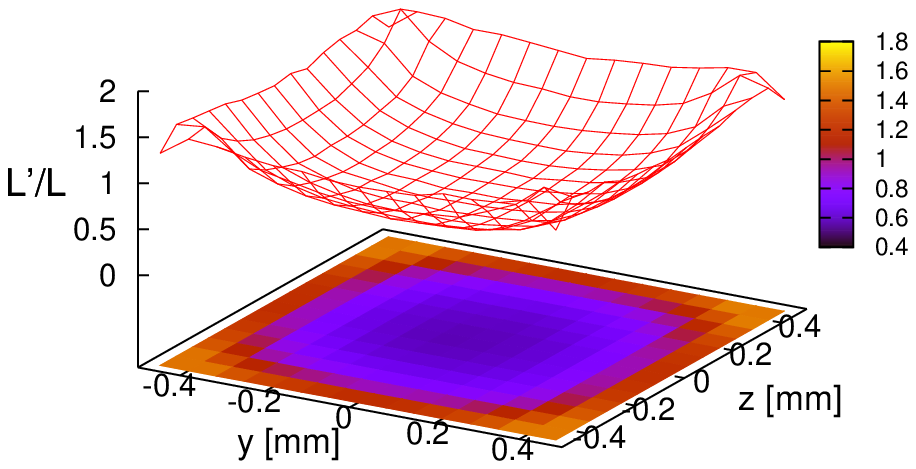}
    (a)
  \end{minipage}
  \hspace{0.01\textwidth}
  \begin{minipage}[b]{0.40\textwidth}
    \includegraphics[width=1.0\textwidth]{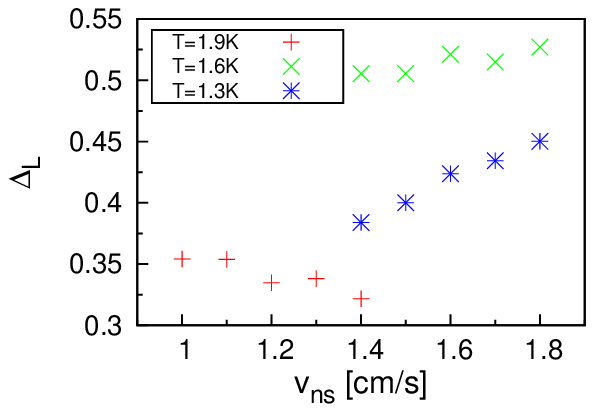}
    (b)
  \end{minipage}
  \end{center}
  \caption
  {
    (Color online) 
    (a) Profiles of the vortex line density ({\it T}=1.9~K, $v_{ns}=1.4{\mathrm{~cm/s}}$).
    One can see the concentration near the solid boundaries.
    (b) Dependence of the spatially variance ${\it \Delta}_L$ of $L'/L$ on $v_{ns}$ and $T$.
  }
  \label{dens.eps}
  \end{figure}

  \section{Discussions}
    Section 4.1 addresses the mechanism for sustaining the inhomogeneous vortex tangle.
    In section 4.2 we discuss how the vortex tangle becomes inhomogeneous depending on temperature.
    In section 4.3 we consider how the aspect ratio of the channel cross section affects the two fluids.

  \subsection{Mechanism for sustaining the inhomogeneous vortex tangle}
    As shown in Fig. 1 (a), the VLD shows the non-linear oscillation with large amplitude in the statistical steady state, which is much different from the case of the uniform counterflow \cite{adachi}.
    The non-linear oscillation comes from the space-time oscillation of the vortex tangle through the mutual friction under the Poiseuille flow.
    Figure 3 shows the space-time pattern of the vortex tangle at 1.9 K.
    The period of the non-linear oscillation is about 0.6 s, consisting of four stages (a)-(d).
    In Fig. 3 (a) corresponding to the minimum of the VLD, vortices are dilute, where vortices remain only near the solid walls.
    Then the vortices near the walls invade to the central region in Fig. 3 (b).
    These vortices make lots of reconnections in the central region subject to the large counterflow in Fig. 3 (c).
    Hence the VLD increases significantly to the maximum.
    Eventually in Fig. 3 (d) the Poiseuille flow excludes the vortex tangle from the central region toward the solid walls.
    Thus the VLD around the central region decreases, and the vortices are absorbed by the solid boundaries.
    Then the vortex tangle returns to the stage of Fig. 3 (a).
    Therefore the vortex tangle repeats the periodic motion, resulting in the non-linear oscillation of Fig. 1 (a).
    The vortex tangle sustained by this mechanism is more dilute than the case of the uniform counterflow for the same $T$ and $v_{ns}$.

    \begin{figure}[b]
    \begin{minipage}[b]{0.245\textwidth}
      \includegraphics[width=1.0\textwidth]{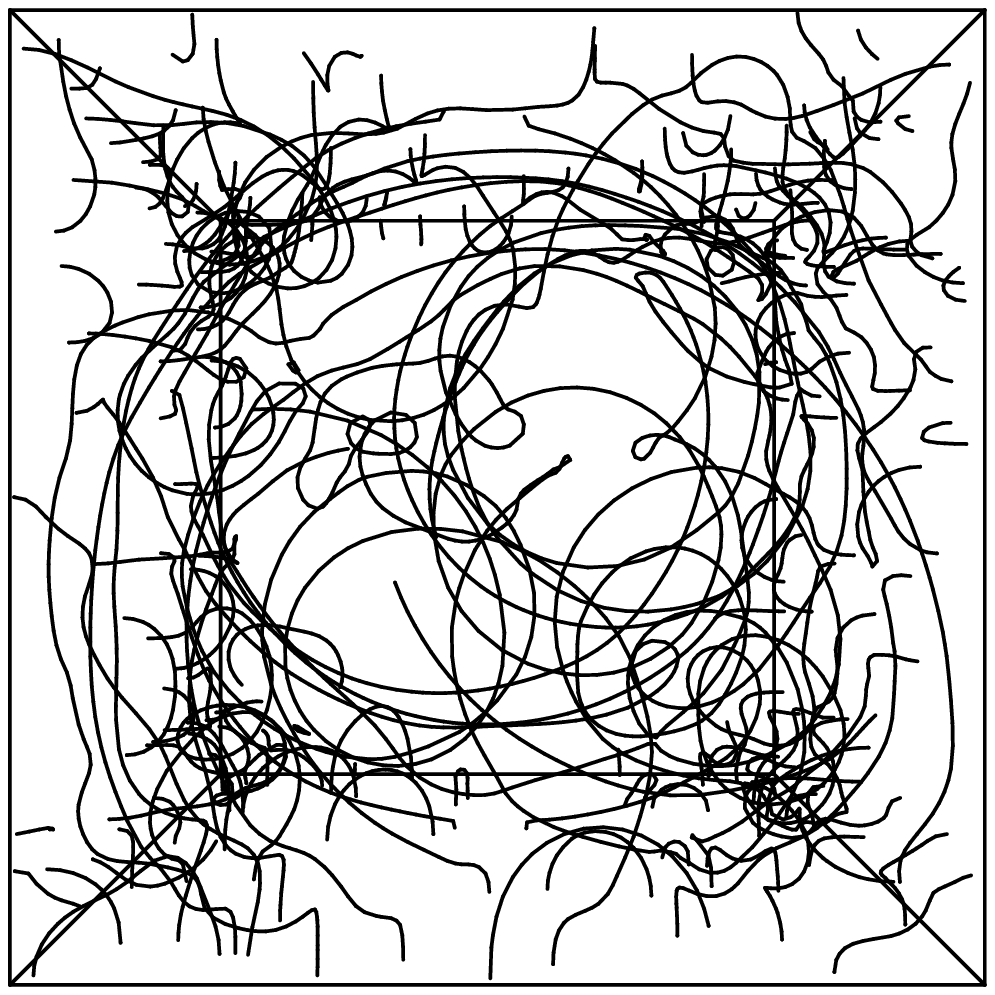}
      (a) Depletion (2.00 s)
    \end{minipage}
    \begin{minipage}[b]{0.245\textwidth}
      \includegraphics[width=1.0\textwidth]{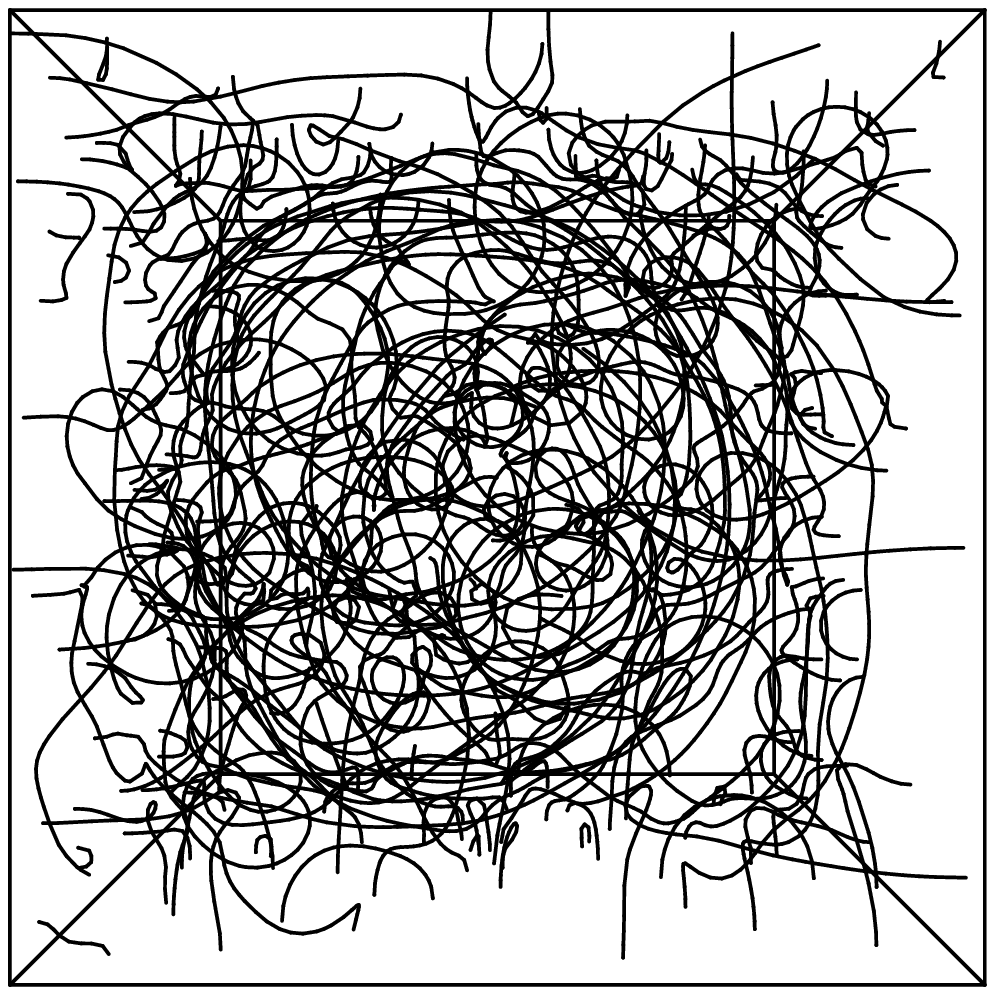}
      (b) Invasion (2.14 s)
    \end{minipage}
    \begin{minipage}[b]{0.245\textwidth}
      \includegraphics[width=1.0\textwidth]{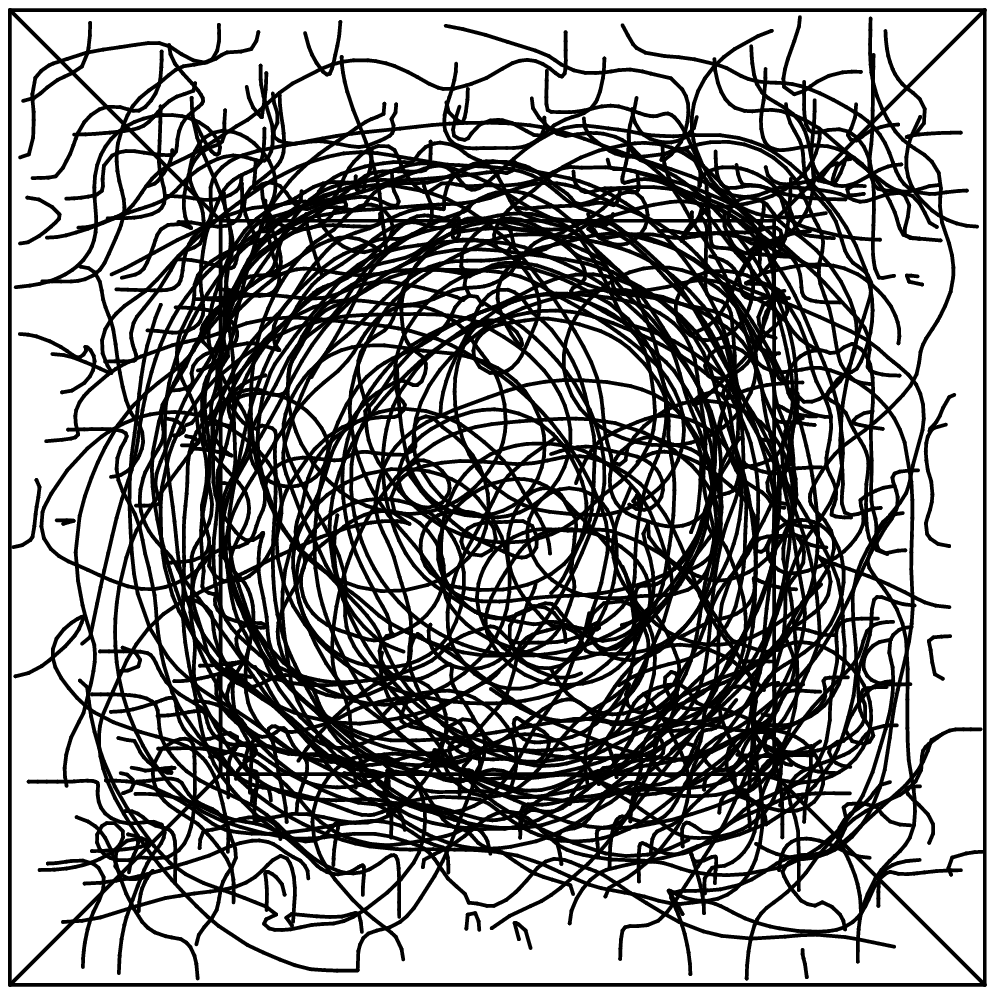}
      (c) Maximum (2.30 s)
    \end{minipage}
    \begin{minipage}[b]{0.245\textwidth}
      \includegraphics[width=1.0\textwidth]{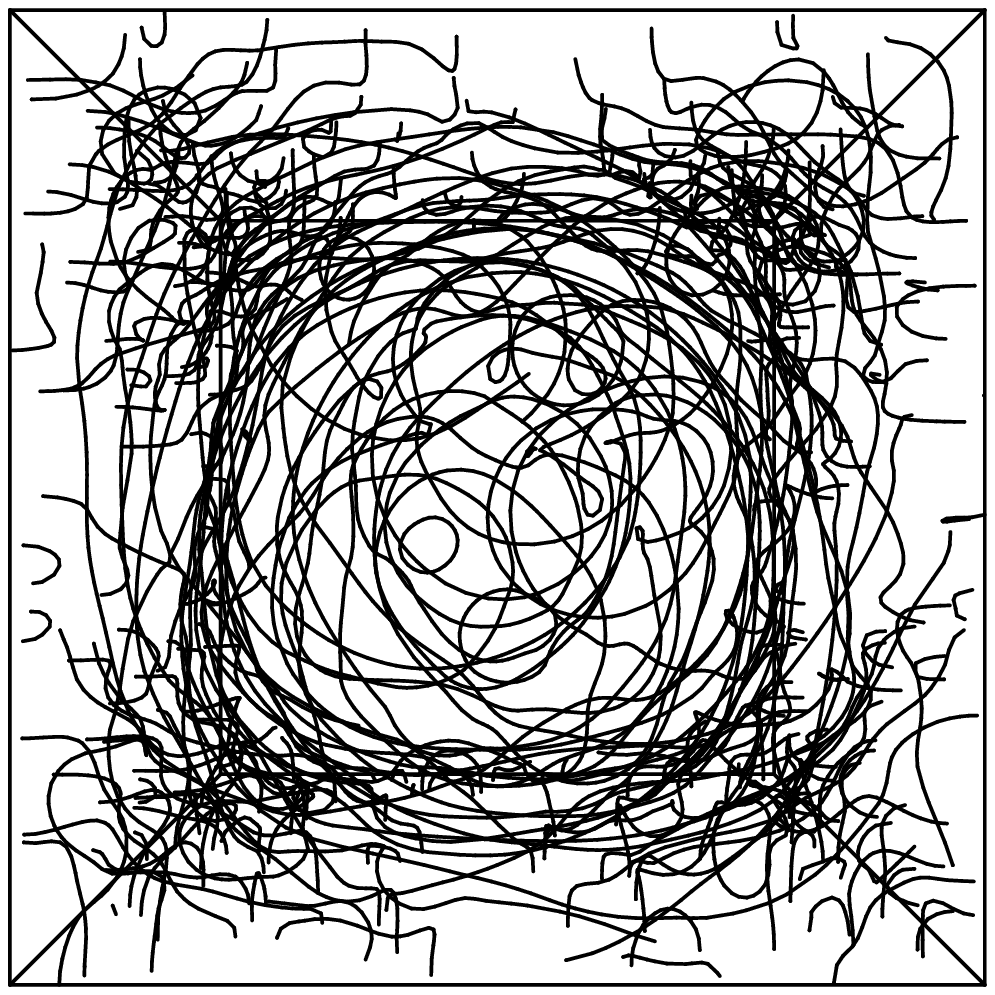}
      (d) Exclusion (2.42 s)
    \end{minipage}
    \caption
    {
      Vortex tangles in the statistically steady state viewed along the flow direction ($T=$1.9~K, $v_{ns}=1.4~{\rm cm}$), corresponding to the results of Fig. 1 (a).
    }
    \label{tangles.eps}
    \end{figure}

  \subsection{Temperature dependence of inhomogeneity}
    Temperature dependence of ${\it \Delta}_L$ in Fig. 2 (b) is caused by the temperature dependent parameters, namely the density ratio $\rho_{ns}=\rho_{n}/\rho_{s}$ of the two fluids and the mutual friction coefficients $\alpha$ and $\alpha'$.
    The increase in temperature causes two competitive effects as described in the following.
    One effect is that the increase of $\rho_{ns}$ decreases $\Delta_L$.
    The thermal counterflow requires the conservation of mass, yielding
    $
      {\bf v}_{s,a} = - \rho_{ns} \overline{{\bf v}}_{n},
    $
    where $\overline{{\bf v}}_n$ is spatially averaged ${\bf v}_n$ over the cross section of the channel.
    The ratio $\rho_{ns}=v_{s,a}/{\overline v}_{n}$ increases with temperature.
    Thus the counterflow velocity ${\bf v}_{ns}={\bf v}_{n} - {\bf v}_{s,a}$ becomes more uniform at higher temperature, since ${\bf v}_{s,a}$ is uniform and ${\bf v}_{n}$ is non-uniform.
    Figure 4 shows the temperature dependence of the counterflow velocity profile.
    The other effect is that the enhancement of the mutual friction increases $\Delta _L$.
    The mutual friction makes the vortex tangle inhomogeneous under the non-uniform ${\bf v}_{ns}$.
    Therefore the increase in temperature renders the profiles of ${\bf v}_{ns}$ uniform but enhances this action of the mutual friction.
    These two competitive effects maximize the inhomogeneity $\Delta _L$ at some $T$.


    \begin{figure}[t]
    \begin{minipage}[b]{0.34\textwidth}
      \includegraphics[width=1.0\textwidth]{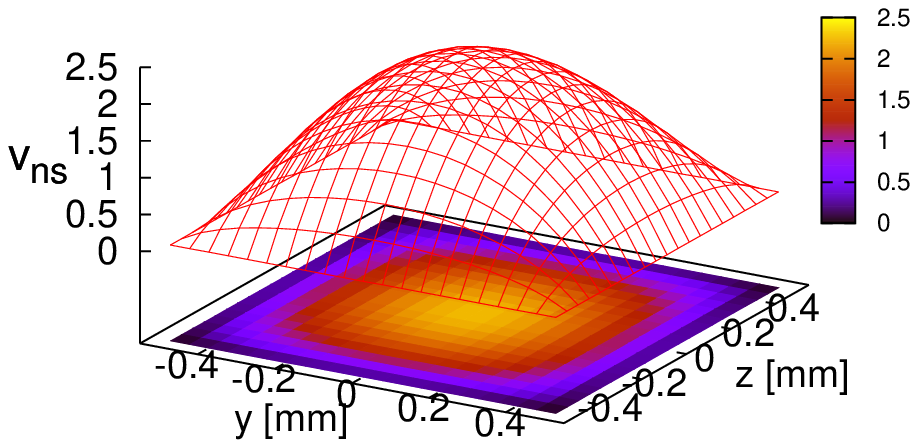}
      (a) $T$=1.3~K ($\rho_{ns}$=0.052)
    \end{minipage}
    \begin{minipage}[b]{0.34\textwidth}
      \includegraphics[width=1.0\textwidth]{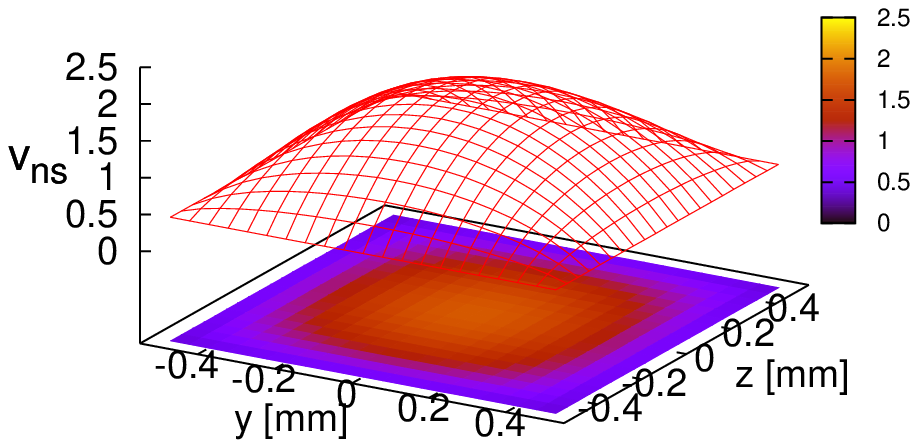}
      (b) $T$=1.9~K ($\rho_{ns}$=0.738)
    \end{minipage}
    \begin{minipage}[b]{0.30\textwidth}
      \caption
        {
          (Color online)
          Counterflow velocity scaled by its spatial average.
        }
      \label{vns.eps}
    \end{minipage}
    \end{figure}

  \subsection{Aspect ratio of the channel cross section}
    The aspect ratio of the channel plays an important role in the density and the profile of the vortex tangle.
    The increase of the aspect ratio from unity reduces the counterflow velocity gradient along the long side of the cross section.
    Thus the exclusion of the vortex tangle shown in Fig. 3 (d) does not work so much along the long side of the cross section.
    Hence the VLD should increase compared to the VLD in a low aspect ratio channel.

    It would be meaningful to consider how the aspect ratio affects the  normal fluid, though it is prescribed in this formulation.
    The linear stability analysis of the Navier-Stokes equation shows that the critical Reynolds number for turbulence transition of a viscous fluid increases significantly with decreasing the aspect ratio \cite{tatsumi}.
    According to this result, the normal fluid in counterflow could remain laminar in a low aspect ratio channel even if the superfluid becomes turbulent and the vortex tangle disturbs the normal fluid, which may correspond to the TI state.
    In order to understand the TI state, therefore, the studies in a low aspect ratio channel like this work will be indispensable.


\ack
  This work was supported by JSPS KAKENHI Grant Number 26400366 and MEXT KAKENHI "Fluctuation \& Structure" Grant Number 26103526. 



\end{document}